\documentclass[runningheads]{llncs}
\usepackage{graphicx}
\usepackage{hyperref}
\usepackage{amsmath}
\usepackage{orcidlink}
\usepackage{multirow}

\usepackage{tikz}
\usetikzlibrary{shapes.geometric}

\begin{document}

\title{Multimodal Embeddings for 3D Similarity Search 
in Semantic Web-of-Things Digital-Twin Platforms}
\titlerunning{Multimodal Embeddings for 3D Digital Twins in SWoT}

\author{Oussama Zaid\inst{1,2}\orcidlink{0009-0001-8770-8700} \and
Romaric Gaudel\inst{1}\orcidlink{0000-0002-6364-5485} \and
Thomas Hassan\inst{2}\orcidlink{0000-0003-2912-1984} \and
Maria Massri\inst{2}\orcidlink{0009-0009-3890-6266} \and
Philippe Raipin-Parvédy\inst{2}\orcidlink{0000-0001-5605-2262}}

\authorrunning{O. Zaid et al.}

\institute{Univ Rennes, Inria, CNRS, IRISA; F-35000 Rennes, France \\
\email{romaric.gaudel@irisa.fr}\and
Orange Research, Cesson-Sévigné, France \\
\email{\{oussama.zaid, thomas.hassan, maria.massri, philippe.raipin\}@orange.com}}

\maketitle

\begin{abstract}
Semantic Web of Things (SWoT) platforms model physical infrastructure as knowledge graphs typed against domain ontologies, enabling expressive structural and logical queries. However, they lack native mechanisms to express similarity beyond strict ontological equivalence, which represents a critical gap for 3D digital twins in domains such as telecom infrastructure and industrial IoT, where queries must combine ontological constraints with multimodal similarity search over heterogeneous, temporally-evolving scene data. We propose a framework that extends SWoT platforms with a multimodal embedding layer: ontology-typed entities comprising 3D point clouds, temporal attributes, and semantic labels are encoded into latent vector representations stored alongside the knowledge graph, enabling hybrid ontology-vector queries that combine graph-based filtering with similarity search. Implemented on Orange Research's Thing'in platform with the Clock-G temporal graph database, a feasibility evaluation on S3DIS demonstrates that graph filtering effectively restricts the search pool under temporal and relational constraints, and that general-purpose pretrained encoders produce representations sufficient for similarity retrieval and as a preliminary encoding step for downstream predictive tasks.
\end{abstract}

\section{Introduction}
Semantic Web of Things (SWoT) platforms can represent physical environments --- such as telecom networks, smart buildings, industrial facilities --- as knowledge graphs where entities are typed against domain ontologies expressed in RDF/OWL \cite{Huang2024}. Vocabularies such as BOT (Building Topology Ontology) \cite{Rasmussen2020} for spatial structures, IFC \cite{IFC2018} for building elements, and SAREF \cite{Daniele2015} for IoT devices provide standardized, machine-readable descriptions that enable interoperability across heterogeneous data sources and semantic querying via SPARQL or Cypher-based languages. With the growing availability of 3D scanning technologies (LiDAR, photogrammetry), these platforms increasingly ingest rich geometric data point clouds, meshes, BIM models alongside their semantic descriptions, giving rise to \textbf{3D digital twins} \cite{Hananto2024,Geng2022} that combine geometric fidelity with ontological context \cite{Ieva2025}. These 3D digital twins already support contextual visualization, property querying, and temporal tracking \cite{Massri2022}. However, current query mechanisms remain fundamentally \textbf{symbolic}: operating on discrete labels and explicit ontological relationships (e.g., ``Which elements belong to Room A?''). They cannot reason about \textbf{continuous similarity}: for instance, ``Which other objects of this type, at this site, during this period, have a 3D shape similar to this one?''. Answering such queries requires transforming heterogeneous WoT data---unstructured 3D point clouds, timestamps, and categorical labels---into compact vector representations (\textit{embeddings}) \cite{Liu2025,Ma2025} that preserve both geometric and semantic properties while remaining integrated with the knowledge graph.
% \oz{à raccourcir, trop détaillé pour une intro?} \mm{(non, pas trop détaillé mais la section état de l'art est un peu longue)}

We propose a \textbf{framework} that extends SWoT platforms by introducing a \textbf{multimodal embedding layer} alongside the existing knowledge graph. Instead of relying solely on symbolic links, each ontology-typed entity is encoded into a unified vector representation capturing its 3D geometry, temporal context, and semantic attributes. These general-purpose representations are indexed by ontological type and timestamp, enabling \textbf{hybrid ontology-vector queries}. This introduces a new reasoning capability: while ontologies provide strict \textit{``is-a''} reasoning (e.g., \texttt{Chair subClassOf Furniture}), the embedding layer enables \textit{``is-similar-to''} reasoning, allowing the system to retrieve entities that share comparable morphological, temporal, and semantic profiles.

The present work is implemented within \textbf{Thing'in} \cite{Raipin2023}, Orange Research's digital twin platform, which models infrastructure as a semantically-grounded property graph \cite{Privat2018}. Thing'in relies on \textbf{Clock-G} \cite{Massri2022}, a temporal graph database in which nodes and edges carry validity timestamps, making it possible to track how both the structure and properties of an environment change over time. We extend this infrastructure with a multimodal embedding pipeline that encodes each entity's 3D geometry, temporal context, and semantic attributes into a unified vector representation, stored alongside the knowledge graph, and introduce a two-phase query mechanism: symbolic graph filtering first isolates entities by ontological type, temporal window, and relational context; vector-based similarity ranking then orders the remaining candidates by proximity in the learned embedding space.
 
We evaluate the hybrid query planner on the S3DIS indoor dataset~\cite{Armeni2017}, modelled as a semantic temporal graph with synthetic snapshots to simulate scene evolution. Two queries are instantiated whose discriminating constraints are inexpressible in the vector index alone: cross-temporal re-identification under a temporal validity constraint, and room-type-scoped similarity search via relational traversal. Results demonstrate that graph filtering is necessary to enforce temporal and relational constraints that the vector index cannot express alone, and that general-purpose pretrained encoders, without task-specific fine-tuning, support effective similarity retrieval and constitute a viable preliminary encoding step for downstream tasks.

The remainder of the paper is organized as follows. Section~2 reviews related work on semantic digital twins, embeddings, and hybrid query architectures. Section~3 introduces the Thing'in platform and Clock-G. Section~4 details the proposed approach. Section~5 presents experimental results, and Section~6 concludes with perspectives for future work.

\section{Related Work}
% \mm{à raccourcir ?}
To contextualize our approach, we review prior work across three intersecting domains that form the foundation of our proposed architecture. First, we examine how knowledge graphs and ontologies have been applied to structure digital twin data. Second, we survey embedding approaches capable of capturing semantic and geometric modalities. Finally, we review the evolution of vector storage and hybrid query architectures.

\begin{enumerate}
\item \textbf{Knowledge Graphs and Ontologies for Digital Twins:}

Graph-based models have become a central paradigm for organizing digital twin data. Huang et al.\ \cite{Huang2024} formalized digital twins as property graphs that manage the complexity and uncertainty of interconnected physical entities. When enriched with domain ontologies, these graphs become semantic knowledge graphs supporting standardized querying and cross-system interoperability. Abbas et al.\ \cite{Privat2018} bridged property graphs and RDF/OWL through blank-node reification, ensuring compatibility via JSON-LD serialization.
 
On the 3D front, Hananto et al.\ \cite{Hananto2024} surveyed the evolution toward immersive 3D digital twins, noting persistent challenges in integrating geometric data with semantic structures. Ieva et al.\ \cite{Ieva2025} recently addressed this gap by annotating 3D indoor scenes with ontology-typed knowledge graphs for smart infrastructure.

The resulting architecture (ontology-typed graphs with 3D geometry attached to described entities) provides powerful symbolic querying capabilities. However, no existing platform offers a mechanism to query entities by the \textit{similarity} of their 3D geometry within the knowledge graph itself.

\vspace{0.3cm}

\item \textbf{Embedding Approaches for Semantic and Geometric Data:}
 
Integrating learned vector representations into graph systems has received growing attention. Besta et al.\ \cite{Besta2022} introduced the concept of \textit{neural graph databases} with LPG2vec, showing that encoding the full property richness of labeled property graphs improved prediction accuracy by up to 34\%. Their work, however, addresses static graphs with textual and scalar properties and does not consider temporal evolution, 3D data, or ontological typing. Within the semantic web community, Chen et al.\ \cite{Chen2021} proposed OWL2Vec*, encoding graph structure together with lexical information and logical constructors of OWL ontologies into vector spaces. While effective for capturing structural proximity, these methods do not encode entity \textit{content} (geometry, sensor readings, or temporal state) that characterizes WoT data.
 
On the geometric side, Bickel et al.\ \cite{Bickel2022} used deep learning embeddings for shape retrieval in mechanical CAD catalogs, and Huang et al.\ \cite{HuangBIM2025} combined geometric, semantic, and topological features for intelligent BIM element search, both confirming the practical value of learned 3D descriptors for asset management.
 
\vspace{0.3cm}

\item \textbf{Vector Storage and Hybrid Query Architectures:}
 
The expansion of vector database capabilities \cite{Ma2025} has created new possibilities for combining structured queries with similarity search. Dedicated systems such as Milvus \cite{Wang2021} and Qdrant \cite{Qdrant2024} offer high-performance approximate nearest-neighbor retrieval, while database extensions like Cassandra~v5's native \texttt{VECTOR} type with SAI indexing \cite{Cassandra2024} and pgvector \cite{pgvector} allow vector capabilities within existing infrastructure. Graph databases such as Neo4j \cite{neo4j2026} and Amazon Neptune \cite{amazon_neptune} have also begun incorporating vector search alongside their traversal engines. In the semantic web community, the combination of structured RDF queries with embedding-based similarity has been explored through systems like Vec2SPARQL \cite{Alshahrani2018}, which extends SPARQL with custom functions for jointly querying structured data and vector representations.

These efforts share a common intuition: graph structure provides precise relational constraints while vector proximity captures soft similarity that logical matching cannot express. Yet none of the existing systems integrates 3D geometric similarity, ontological typing, and temporal historization within a single query framework: the gap our work addresses.
\end{enumerate}

\section{Background}

Our framework extends an existing SWoT platform with a vector embedding layer. To situate the contribution precisely, we first describe the platform's data model and querying mechanisms. We focus on Thing'in, a production-grade digital twin platform, and its associated temporal graph database Clock-G.

Thing in the Future, \textit{referred to as "Thing'in" in this paper}, is a digital twin platform developed at Orange Research for modeling and managing complex physical environments such as telecom networks, smart buildings, smart cities, and industrial facilities \cite{Raipin2023}. The platform has been deployed across multiple use cases at Orange, including telecom network supervision, BIM-based building management, and augmented technician assistance.
The data model is built on a \textbf{property graph} storage layer, where nodes represent physical entities (antennas, rooms, sensors, building elements) and edges encode their relationships (containment, connectivity, ownership). Each entity carries a set of typed key-value properties describing its attributes. Critically, this property graph is not a purely structural store: it is given formal semantic grounding through RDF/OWL. Entities are typed against domain ontologies via class URIs;  the platform is not restricted to a fixed vocabulary, allowing operators to use or combine existing ontologies. The platform supports SPARQL querying, and provides data injectors that ingest heterogeneous sources into the unified graph representation.
Thing'in already supports the visualization of 3D assets (point clouds, meshes) alongside their graph representation: an object's geometry can be rendered in a 3D viewer while its ontological properties and relationships are displayed from the knowledge graph. This combination of geometric and semantic data makes the platform a natural candidate for the kind of hybrid querying we propose.
% \mm{The figure is too large and the text is hard to read. I'd rather have a figure in the intro that captures the big picture of this work}

% \pr{transition}
To manage the temporal evolution of these environments, the platform relies on a dedicated storage backend. \textbf{Clock-G} is a temporal graph database, also developed at Orange Research, designed for efficient historization and temporal querying \cite{Massri2022}. In such a temporal graph, nodes, edges, and their attributes are associated with timestamps or validity intervals, allowing the system to track how both structure and properties change over time. Built on Apache Cassandra, Clock-G implements a $\delta$-Copy+Log storage strategy that combines periodic full snapshots with incremental change logs, minimizing storage overhead while maintaining fast access to any historical state. It extends the Cypher query language into \textbf{T-Cypher}, which adds temporal operators for point-in-time queries (retrieving the graph state at a given instant), interval queries (retrieving entities valid during a period), and progressive temporal traversal.

\vspace{0.3cm}

In this work, we extend this infrastructure by adding a vector embedding layer on top of the existing semantic and temporal layers, enabling hybrid queries that combine Thing'in's ontological filtering and Clock-G's temporal constraints with similarity search over learned 3D representations.

\section{Proposed Approach}
 
\begin{figure*}[htbp]
    \includegraphics[width=\textwidth]{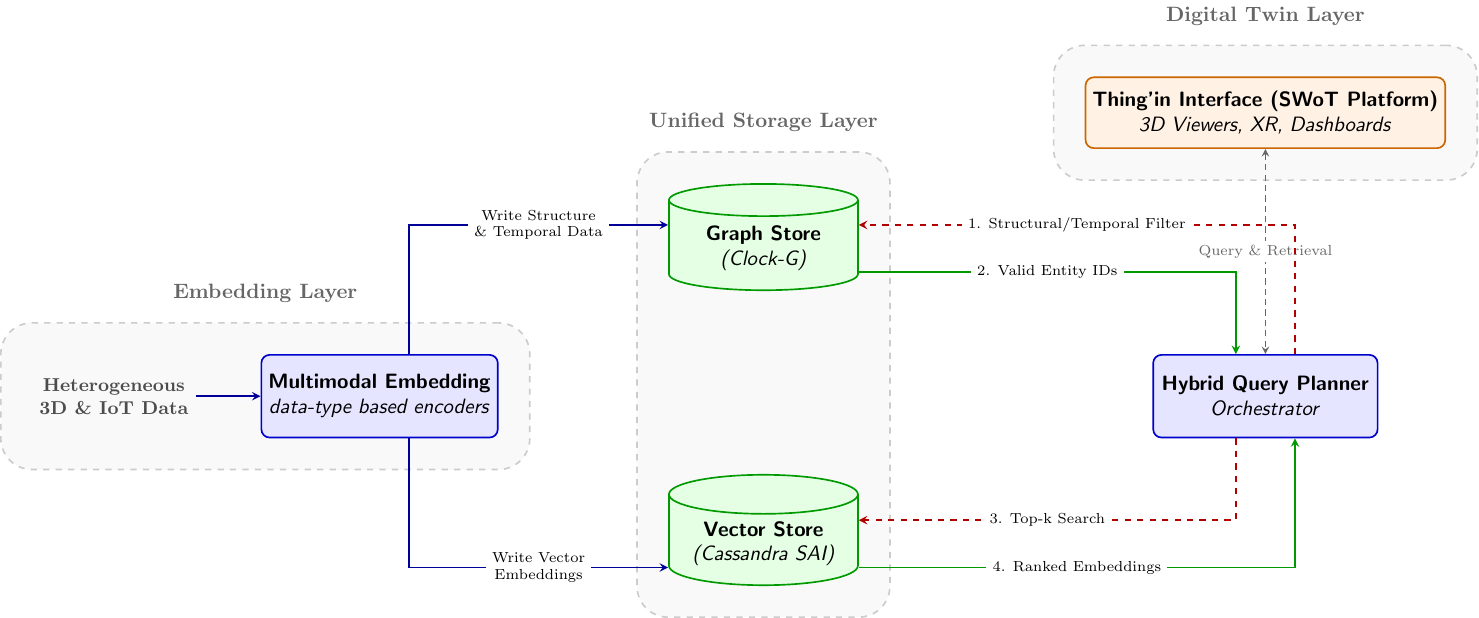}
    \caption{Architecture overview. Heterogeneous entity data flows through three layers: encoding, unified graph-vector storage, and a two-phase hybrid query planner combining symbolic filtering with similarity ranking.}
    \label{fig:global_architecture}
\end{figure*}

\autoref{fig:global_architecture} presents the proposed architecture, organized into three layers: an \textit{Embedding Layer} that transforms heterogeneous entity data into vector representations, a \textit{Unified Storage Layer} that co-locates the temporal knowledge graph with the generated embeddings, and a \textit{Hybrid Query Planner} that coordinates retrieval across both stores. The Thing'in platform operates at the application level, exposing results through its existing interfaces. We detail each layer below.

\subsection{Embedding Layer}
 
The embedding layer receives the heterogeneous properties associated with each ontology-typed entity in the temporal graph and produces a single vector per entity using modality-specific encoders. A central design requirement is ontological coherence: entities sharing a common OWL superclass should yield closer representations than entities from disjoint branches of the class hierarchy.

\textit{Categorical attributes.} Each property-value pair (e.g., \texttt{object\_class: Chair}, where the value denotes an OWL class) is formulated as a natural-language sentence and encoded by Sentence-BERT~\cite{Reimers2019}, which produces a dense vector capturing the semantics of both the ontology property and its associated class. This strategy generalizes to previously unseen ontology concepts without retraining, a necessary property for open-vocabulary SWoT platforms.

\textit{Numerical and temporal attributes.} Scalar measurements and timestamps are projected into a multi-frequency sinusoidal space using Fourier feature mappings~\cite{Tancik2020}. This encoding preserves periodic regularities, such as diurnal or seasonal patterns in infrastructure activity, that flat normalization would suppress. Timestamps are further augmented with derived cyclical components (day of week, hour of day) prior to projection.
 
\textit{3D point clouds.} Each point cloud is centered, normalized to unit scale, and resampled to a fixed cardinality. A pretrained PointNet~\cite{Qi2017} backbone then extracts a global shape descriptor invariant to point ordering, capturing the morphological signature of the physical entity independently of acquisition conditions.
 
\textit{Aggregation.} The per-property vectors are stacked and summarized through statistical pooling (mean, variance, max), then projected to a fixed dimensionality and $L_2$-normalized. This yields one embedding per entity per timestamp, irrespective of the number or nature of properties exposed, a property that accommodates the structural heterogeneity inherent in SWoT knowledge graphs.
 
\subsection{Unified Storage Layer}
The Unified Storage Layer relies on two stores that share a common Apache Cassandra backend. The \textbf{Graph Store} is managed by Clock-G, which historizes the knowledge graph as periodic full snapshots. These are complemented by incremental change logs, allowing any past state of the graph to be reconstructed efficiently. The \textbf{Vector Store} holds the embeddings produced by the encoding layer, indexed via Cassandra~v5's native \texttt{VECTOR} data type and Storage-Attached Indexing (SAI)~\cite{Cassandra2024}, which supports cosine-similarity search without requiring an external vector database.
 
The embedding table is partitioned by \textbf{OWL class URI} and encoding-model version, with timestamp and entity identifier as clustering columns. A similarity query over \texttt{bot:Element} instances will never compare against \texttt{saref:Device} embeddings, as they reside in disjoint partitions. Within a partition, three access patterns are natively supported: retrieval of the most recent embeddings for a given type, temporal trajectory scanning for a specific entity, and similarity search restricted to a temporal window.
 
\subsection{Hybrid Query Planner}
 
\begin{figure}[htbp]

    \centering

    \includegraphics[width=0.86\textwidth]{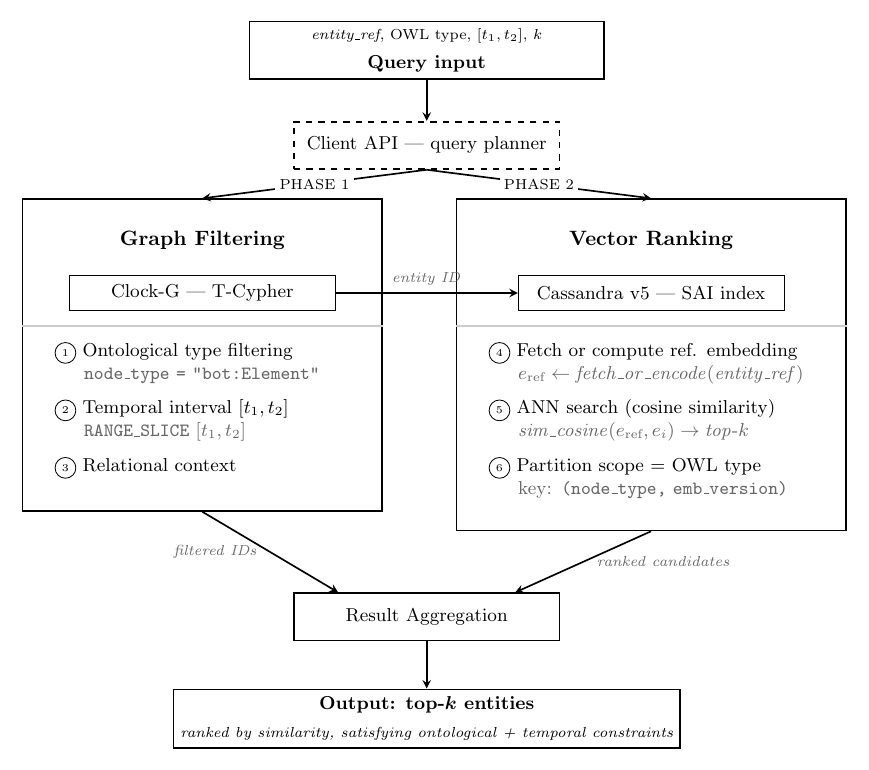}

    \caption{Two-phase query execution architecture. The system decomposes hybrid queries into (1) Graph Filtering via Clock-G, which resolves strict ontological and temporal constraints, and (2) Vector Ranking via Cassandra's SAI index, which fetches or computes the reference embedding to perform an ANN search. The results are aggregated to return the top-k structurally and semantically valid entities.}

    \label{fig:hybrid_query_architecture}

\end{figure}

The query planner coordinates hybrid retrieval across both stores in two sequential phases (\autoref{fig:hybrid_query_architecture}), given a reference entity, an OWL type constraint, a temporal interval, and a neighbor count~$k$.

\paragraph{Phase~1: Graph filtering.} The planner issues a T-Cypher query to Clock-G, applying three successive constraints:
\begin{enumerate}
    \item \textit{Ontological type filtering}: restricts candidates to entities matching the specified OWL class;
    \item \textit{Temporal slicing}: retains only entities whose validity interval intersects the requested period;
    \item \textit{Relational traversal}: follows semantic relationships to further narrow the candidate set according to structural context.
\end{enumerate}
The graph store returns the qualifying entity identifiers along with the reference entity's embedding, which are forwarded to the second phase.
 
\paragraph{Phase~2: Vector ranking.} The planner queries the Cassandra SAI index in three steps:
\begin{enumerate}
    \setcounter{enumi}{3}
    \item \textit{Reference retrieval}: the embedding of the query entity is fetched from the vector store if available, or computed on the fly using the encoding pipeline described in Section~4.1;
    \item \textit{Approximate nearest-neighbor search}: candidate embeddings are ranked by cosine similarity against the reference, returning the top-$k$ results;
    \item \textit{Partitioned search}: the search is confined to the partition corresponding to the target OWL class and encoder version (the embedding pipeline used), ensuring type-consistent comparisons.
\end{enumerate}
 
The planner merges the outputs of both phases: the ranked list from the vector store is intersected with the filtered identifiers from the graph store, yielding a final set of top-$k$ entities that are simultaneously ontologically valid, temporally consistent, and geometrically proximate to the reference.

\section{Experimental Evaluation}

We evaluate the hybrid query planner end-to-end, asking whether combining symbolic graph filtering with vector-based ranking produces retrieval behaviour that neither store could deliver alone.  We instantiate two queries whose discriminating constraints are inexpressible in the vector index in isolation: Q1 imposes temporal validity, and Q2 imposes relational containment. 

\subsection{Setup}

We use the S3DIS indoor dataset~\cite{Armeni2017}, modelled as a semantic temporal knowledge graph in Clock-G across three synthetic snapshots: the original scene~(A), a re-layout with rotations and
partial occlusions~(B), and a renovation with object replacements and removals~(C).  The vector store holds 7\,833 embeddings at dimension 512 across all snapshots, pre-computed using our multimodal embedding pipeline.  The candidate space per snapshot is approximately 2\,600 objects spanning 13 semantic classes across 171 rooms.  Queries are evaluated against this full data using the two-phase planner described in Section~4.3.

\subsection{Q1 — Cross-Temporal Re-identification}

The purpose of this query is to evaluate whether the pipeline can successfully re-identify the same physical entity across temporal snapshots. Given a reference entity $x_0$ at a source snapshot, Phase 1 retrieves candidate nodes from the target snapshot via a Clock-G T-Cypher request constrained to the target time window. In Phase 2, vector similarity is computed over the retrieved candidates to identify the closest match to $x_0$. Ground truth is entity-ID continuity; and 300 references are sampled per directional pair from the 2\,475 entities persistent across all three snapshots.

Table~\ref{tab:q1} reports results for the three directional pairs.
Performance degrades with transformation severity: A$\to$B achieves Hit@1~=~0.910, while A$\to$C drops to 0.837 which is geometrically expected, since renovation replaces point cloud geometry entirely.  B$\to$C (0.900) is close to A$\to$B because it constitutes a single mild transformation; degradation is driven by transformation
\emph{type}, not cumulative temporal distance.  Recall@10 on the hardest condition remains 0.957, confirming the embedding space retains
value as a re-identification prior even when rank-1 precision fails. 
\begin{table}[h]
\centering
\caption{Q1 — Cross-temporal re-identification. $N=300$ per condition. Values are mean\,$\pm$\,std over per-query observations.}
\label{tab:q1}
\begin{tabular}{lccccc}
\hline
Condition & Hit@1\,\textsubscript{ANN} & Hit@1\,$\uparrow$ & MRR\,$\uparrow$ & R@5\,$\uparrow$ & R@10\,$\uparrow$ \\
\hline
A$\to$B (re-layout)   & 0.907 & 0.910\,{\scriptsize$\pm$0.287} & 0.934\,{\scriptsize$\pm$0.220} & 0.963\,{\scriptsize$\pm$0.188} & 0.973\,{\scriptsize$\pm$0.161} \\
A$\to$C (renovation)  & 0.473 & 0.837\,{\scriptsize$\pm$0.370} & 0.875\,{\scriptsize$\pm$0.294} & 0.920\,{\scriptsize$\pm$0.272} & 0.957\,{\scriptsize$\pm$0.204} \\
B$\to$C (incremental) & 0.520 & 0.900\,{\scriptsize$\pm$0.301} & 0.930\,{\scriptsize$\pm$0.217} & 0.970\,{\scriptsize$\pm$0.171} & 0.983\,{\scriptsize$\pm$0.128} \\
\hline
\end{tabular}
\end{table}

\subsection{Q2 — Room-Type-Scoped Similarity}

The purpose of this query is to evaluate whether the pipeline can successfully 
combine graph traversal with vector similarity search. Given a reference entity 
of class $c$ at snapshot $s$ and a room-type label $\tau \in \{\textit{office}, \textit{hallway}, \textit{auditorium}\}$, Phase~1 resolves the spatial scope by traversing \texttt{bot:containsElement} edges to retrieve all candidate nodes contained in rooms of type $\tau$ at $s$. In Phase~2, vector similarity is  computed over the retrieved candidates to identify the top-$k$ most similar objects to the reference entity. Ground truth is same-class membership within  the scoped pool; $N = 100$ references per room type per snapshot.

Table~\ref{tab:q2} reports results. Phase~1 reduces the search space by 53\%, 76\%, and 87\% for offices, auditoriums, and hallways respectively, confirming substantial constraint work. P@1 and P@5 are near-ceiling but partly reflect class distribution within room types rather than fine-grained discriminative power; P@10 (0.861--0.900) is the more informative metric, exposing the soft class boundaries of the embedding space. Results remain stable across snapshots despite growing pool sizes.

\begin{table}[h]
\centering
\caption{Q2 — Room-type-scoped similarity. 100 references sampled per (room type, snapshot) pair; pool = candidate set size after graph filtering. Values are mean\,$\pm$\,std over per-query observations.}
\label{tab:q2}
\setlength{\tabcolsep}{4pt}
\begin{tabular}{llcccc c}
\hline
Room type & Snapshot & P@1\,$\uparrow$ & P@5\,$\uparrow$ & P@10\,\textsubscript{ANN} & P@10\,$\uparrow$ & Pool \\
\hline
\multirow{3}{*}{Office}
 & A & 1.000\,{\scriptsize$\pm$0.000} & 1.000\,{\scriptsize$\pm$0.000} & 0.980 & 0.900\,{\scriptsize$\pm$0.000} & 1{,}216 \\
 & B & 1.000\,{\scriptsize$\pm$0.000} & 1.000\,{\scriptsize$\pm$0.000} & 0.670 & 0.900\,{\scriptsize$\pm$0.000} & 1{,}320 \\
 & C & 1.000\,{\scriptsize$\pm$0.000} & 1.000\,{\scriptsize$\pm$0.000} & 0.912 & 0.900\,{\scriptsize$\pm$0.000} & 1{,}382 \\
\hline
\multirow{3}{*}{Hallway}
 & A & 1.000\,{\scriptsize$\pm$0.000} & 0.994\,{\scriptsize$\pm$0.060} & 0.977 & 0.861\,{\scriptsize$\pm$0.116} & 329 \\
 & B & 1.000\,{\scriptsize$\pm$0.000} & 0.994\,{\scriptsize$\pm$0.060} & 0.687 & 0.882\,{\scriptsize$\pm$0.094} & 344 \\
 & C & 1.000\,{\scriptsize$\pm$0.000} & 0.994\,{\scriptsize$\pm$0.060} & 0.976 & 0.885\,{\scriptsize$\pm$0.089} & 366 \\
\hline
\multirow{3}{*}{Auditorium}
 & A & 1.000\,{\scriptsize$\pm$0.000} & 0.994\,{\scriptsize$\pm$0.045} & 0.990 & 0.879\,{\scriptsize$\pm$0.096} & 615 \\
 & B & 0.990\,{\scriptsize$\pm$0.100} & 0.990\,{\scriptsize$\pm$0.100} & 0.970 & 0.882\,{\scriptsize$\pm$0.106} & 623 \\
 & C & 0.990\,{\scriptsize$\pm$0.100} & 0.990\,{\scriptsize$\pm$0.100} & 0.995 & 0.886\,{\scriptsize$\pm$0.098} & 627 \\
\hline
\end{tabular}
\end{table}

\subsection{Comparison with ANN-Only Retrieval}

Tables~\ref{tab:q1} and~\ref{tab:q2} include results for an ANN-only baseline that searches the full per-snapshot corpus without graph
filtering.  For Q1, the baseline performs comparably on A$\to$B (Hit@1~=~0.907) but collapses on A$\to$C and B$\to$C (0.473 and
0.520 respectively), since without a temporal constraint it retrieves embeddings indiscriminately across snapshots.  The graph layer is
therefore not an optimisation for Q1 — it is what makes the query temporally meaningful.  For Q2, the baseline exhibits large snapshot-dependent variance in P@10 (dropping to 0.670 for office and
0.687 for hallway at snapshot~B), while the hybrid system remains stable across all conditions.  In conditions where ANN-only P@10 exceeds the hybrid, the gain comes from retrieving same-class objects from other snapshots or room types, trading temporal and relational correctness for apparent precision.

% \vspace{-2.6mm}

\subsection{Discussion}
The results support three claims. First, the hybrid architecture is necessary: temporal scoping in Q1 and relational containment in Q2 are constraints the vector index cannot express alone, while the graph layer cannot rank by similarity — neither component is sufficient single-handedly in structural retrieval use-cases. Second, without any task-specific training, a general-purpose embedding pipeline is a viable and practical foundation for digital twin platforms, where ML objectives are rarely defined upfront; encoding entities before any learning objective exists allows the platform to support similarity search immediately and positions it for downstream tasks once objectives emerge. Third, the pipeline behaves predictably under scene evolution: re-identification holds under mild geometric perturbation (Hit@1 $\geq$ 0.90) and degrades gracefully under structural renovation, which is precisely the failure mode a platform operator needs to understand before relying on similarity-based reasoning in production.
\enlargethispage{3\baselineskip}
\section{Conclusion and Future Work}
 
This paper presented a framework for extending Semantic Web of Things platforms with a multimodal embedding layer that enables similarity-based reasoning alongside traditional ontological querying. The framework introduces three components: modality-specific encoders that project heterogeneous entity properties into a unified vector space; a storage design that co-locates these vectors with the temporal knowledge graph on a shared backend; and a hybrid query planner that chains symbolic graph filtering with vector-based nearest-neighbor ranking, implemented on Orange Research's Thing'in platform and its temporal graph database, Clock-G.
 
Experimental evaluation on the S3DIS dataset, modelled as a semantic temporal graph with synthetic temporal snapshots, assessed the hybrid query planner end-to-end through two queries requiring constraints
inexpressible in the vector index alone.  Results indicate that preliminary graph filtering meaningfully restricts the search pool and is necessary when temporal validity must be enforced exactly.  They
further show that general-purpose pretrained encoders, without task-specific fine-tuning, produce representations that support effective similarity retrieval and constitute a viable preliminary encoding step for downstream tasks.

Several directions remain open. The current embeddings are produced independently per entity; graph neural networks (GNNs) could refine them by propagating relational context across the knowledge graph. The evaluation relies on synthetic temporal evolution and requires validation on real-world Thing'in deployments such as Orange telecom site monitoring or BIM-based building management. Extending T-Cypher --- Clock-G's temporal graph query language --- with native similarity operators would enable declarative hybrid queries. Finally, the embeddings are designed as input features for downstream predictive tasks (anomaly detection, change forecasting), a direction we intend to explore within the proposed framework.

\bibliographystyle{splncs04} 
\bibliography{references}    
\end{document}